\documentclass{aastex63}

\graphicspath{{./}{figures/}}
\begin{document}

\title{An Analysis of X-Ray Hardness Ratios Between Asynchronous and Non-Asynchronous Polars}

\correspondingauthor{Eric Masington}
\email{eric.masington@ttu.edu}

\author{Eric Masington}
\altaffiliation{Student}
\affiliation{Department of Physics and Astronomy, Texas Tech University, Lubbock, TX 79409, USA}

\author{Thomas J. Maccarone}
\affiliation{Department of Physics and Astronomy, Texas Tech University, Lubbock, TX 79409, USA}

\author{Liliana Rivera Sandoval}
\affiliation{Department of Physics and Astronomy, Texas Tech University, Lubbock, TX 79409, USA}
\affiliation{Department of Physics, University of Alberta, Edmonton, Alberta T6G 2R3, CA}

\author{Craig Heinke}
\affiliation{Department of Physics, University of Alberta, Edmonton, Alberta T6G 2R3, CA}

\author{Arash Bahramian}
\affiliation{Department of Physics and Astronomy, Curtin University, Perth, Western Australia 6845, AU}

\author{Aarran W. Shaw}
\affiliation{Department of Physics, University of Nevada, Reno, NV 89557, USA}

\begin{abstract}

The subclass of magnetic Cataclysmic Variables (CV), known as asynchronous polars, are still relatively poorly understood. An asynchronous polar is a polar in which the spin period of the white dwarf is either shorter or longer than the binary orbital period (typically within a few percent). The asynchronous polars have been disproportionately detected in soft gamma-ray observations, leading us to consider the possibility that they have intrinsically harder X-ray spectra. We compared standard and asynchronous polars in order to examine the relationship between a CV's synchronization status and its spectral shape.  Using the entire sample of asynchronous polars, we find that the asynchronous polars may, indeed, have harder spectra, but that the result is not statistically significant. 
\end{abstract}



\section{Introduction}

One of the first results on accreting white dwarfs with the International Gamma-Ray Astrophysical Laboratory (INTEGRAL) was that the asynchronous polars represented a disproportionate fraction of its detected cataclysmic variables \citep{2006MNRAS.372..224B}.  Asynchronous polars are magnetically accreting white dwarfs with deviations between the orbital and spin period (unlike standard polars) and with streamlike accretion rather than accretion disks (unlike the intermediate polars).  The difference between the spin and orbital period in the asynchronous polars is typically about 5\% or less.  It is unclear whether INTEGRAL preferentially detected these objects because the X-ray and $\gamma-$ray spectra of the asynchronous polars are different from those of the standard polars, or merely because they tend to be more luminous, hence at higher fluxes within the well-understood samples.  Here, we test whether the spectral indices of these sources in the soft $\gamma-$ray band alone, and between X-ray and $\gamma$-ray, are systematically harder for the asynchronous polars than for the standard polars.

The asynchronous polars are often suggested to have been driven out of synchronization by classical novae, which can affect both the orbital and spin periods of cataclysmic variables, motivated by the association between one of the asynchronous polars, V1500~Cyg with a classical nova in 1975\citep{1999A&A...343..132C}.  Searches for additional nova shells around other asynchronous polars have not yielded any new evidence for the nova hypothesis \citep{2016MNRAS.458.1833P}, but it remains a viable one, as nova shells may have lifetimes shorter than the synchronization timescales of the asynchronous polars.  Because the sample sizes of the asynchronous polars are quite small, and rather long, well-sampled light curves are needed to identify that there are two separate, but similar periods in the light curves, it is worth exploring new methods that might work to find new members of the class, and the INTEGRAL discoveries of these objects suggest that gamma-ray surveys might be an interesting approach.  With this in mind, we undertake an exploration of whether the high energy spectra of asynchronous polars are fundamentally different from those of the standard polars.  

\section{Data used}

We obtained a set of cataclysmic variables from the Ritter and Kolb catalog, update 7.24 \citep{2003AA&A.404..301}, hard X-ray data from the 2018 Swift-BAT 105-month All-Sky Survey catalogue \citep{2018AAS.235..1} and soft X-ray data from the ROSAT All-Sky Survey \citep{2016A&A.588..A103}. Matching all three catalogues with a maximum separation of 3 arc minutes (typical for Swift-BAT for faint sources) yields 10 objects, including 4 asynchronous polars and 6 standard polars. We define an asynchronous polar to be a system with a spin period within 5\% of the orbital period.  The asynchronous polars are BY~Cam, CD~Ind, V1432 Aql, and the recently identified IGR~J19552+0044 \footnote{We consider a 5 sigma upper limit for V1500~Cyg and Rx J0838.7-2827 based on the survey depth, but do not consider upper limits for the much larger class of synchronous polars.} \citep{2017AA...608...A36}. V1500~Cyg and RX J0838.7-2827 were added to the analysis to provide for a more complete sample. The synchronous polars included are AM~Her, Swift~J231930.4+261517, 1RXS~145341.1-552146, IW~Eri, V2301~Oph, and V834~Cen. 

\begin{figure}[h!]
\begin{center}
\includegraphics[scale=1,angle=0]{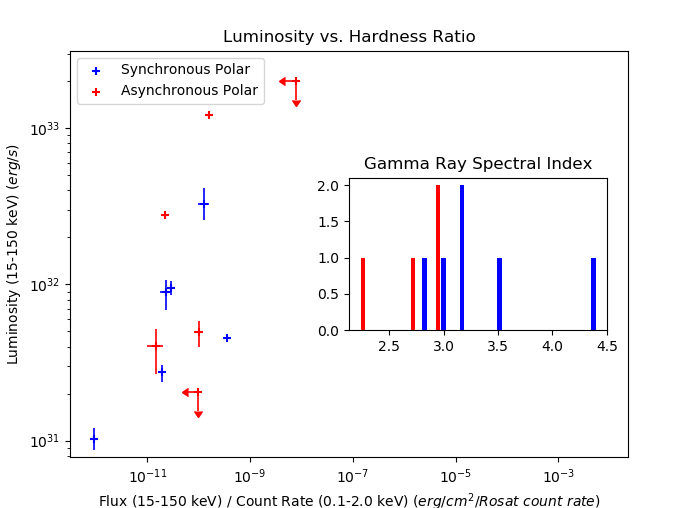}
\caption{Shown here is the ratio between the flux (Swift-BAT 15-150 keV band) and count rate (ROSAT 0.1-2.0 keV band) plotted against luminosity (calculated using {\it Gaia} distances and {\it Swift} flux values). As can be seen, no significant relationship can be established between whether or not a polar is asynchronous and the hardness of its flux ratio. It must be considered that the ROSAT and Swift data are taken non-simultaneously, so because the ROSAT count numbers are generally too small for spectral fitting, they cannot be reliably converted into fluxes. Thus, this means that only very strong trends could have been detected using this combination of data.  Such trends are not present, but the method would not be particularly sensitive to subtle systematic variations. Also shown are the gamma-ray spectral indices. We can see that the asynchronous polars seem to be harder (lower spectral indices), but more data is needed to confirm this. \citep{2018A&A...616A...1G} 
\label{fig:1}}
\end{center}
\end{figure}

\section{Analysis and conclusions}

First, we looked at the ratio of hard X-rays from Swift-BAT (15-150 keV), to soft X-rays from ROSAT (0.1-2.0 keV). This comparison is done between a count rate for ROSAT and a flux for Swift BAT because the standard ROSAT data include only a count rate, and the standard BAT data include only a flux.  The fluxes do, thus, show some model dependence, but since the spectra are all steep power laws, with photon index greater than 2.0, in all cases, the BAT flux is dominated by counts near the lower end of the band, and this comparison is nearly equivalent to a count rate-to-count rate comparison. However,  the ROSAT and Swift data are taken non-simultaneously, and because the ROSAT count numbers are generally too small for spectral fitting, they cannot be reliably converted into fluxes.  This thus means that only very strong trends could have been detected using this combination of data.  Such trends are not present, but the method would not be particularly sensitive to subtle systematic variations. No trend is found in this ratio between the two classes of polars. Also, there exists much uncertainty in the soft X-ray flux values for V1500~Cyg and RX J0838.7-2827, given that they are not in the Swift catalog.  Because the plot shows strong scatter between the Swift and ROSAT data, and the ROSAT data in most cases are insufficient for detailed spectral analysis, we simply leave this plot as a ratio of a flux to a count rate.

Next, we consider the spectra within the $\gamma-$ray band alone. Within the Swift band, the mean spectral index is 2.73 for the asynchronous polars and 3.34 for the standard polars.  We apply the Anderson-Darling test to the distributions of spectral indices.  This is a cumulative statistic, similar to the Kolomogorov-Smirnov test, but with greater diagnostic power in cases where the differences are strongest near the edges of the distributions, and at least equal power in all cases. The Anderson-Darling (AD) test statistic here is 2.28 (computed using https://www.real-statistics.com/non-parametric-tests/goodness-of-fit-tests/two-sample-anderson-darling-test/).  For this sample size, the critical value of the AD test statistic is 3.38 for a 99\% confidence level detection of a difference.  The asynchronous polars do show a different $\gamma$-ray spectral index at the 95\% confidence level.  Since this is a marginally significant difference, we expect that a larger sample of objects would have a reasonable probability of establishing a difference. Unfortunately, doing so will require {\it finding} new asynchronous polars, as we have already investigated the properties of the whole sample, with only V1500~Cyg and RX J0838.7-2827 undetected in the Swift-BAT data.  If more asynchronous polars can be found from optical or soft X-ray searches, NuSTAR would easily be capable of measuring $\gamma$-ray spectral indices for objects much fainter than the BAT survey can, so searches for more asynchronous polars would be well-motivated.

\end{document}